\documentclass[9pt,twocolumn]{article}
\pdfoutput=1
\usepackage[utf8x]{inputenc}
\usepackage{geometry}
\usepackage[subtle]{savetrees}
\usepackage{libertine}
\usepackage{zi4}

\usepackage[T1]{fontenc}
\usepackage{subcaption}
\usepackage{listings}
\usepackage[svgnames]{xcolor}
\usepackage{pgf}
\usepackage{tikz}
\usepackage{tikzscale}
\usepackage{pgfplots}
\usepackage{xspace}
\usepackage{hyperref}
\definecolor{ColorForLink}{named}{DarkRed}
\definecolor{ColorForCite}{named}{DarkOliveGreen}
\definecolor{URLLinkColor}{named}{MidnightBlue}

\hypersetup{
    pdftitle={High-Performance Code Generation though Fusion and Vectorization},
    pdfauthor={J. Sewall and S.J. Pennycook},
    linkcolor=ColorForLink,
    citecolor=ColorForCite,
    urlcolor=URLLinkColor
  }

\usepackage[textsize=tiny,disable]{todonotes}

\definecolor{cinnamon}{rgb}{0.82, 0.41, 0.12}
\definecolor{darkgreen}{rgb}{0.2, 0.41, 0.3}

\graphicspath{{figures/}}

\newcommand{\eg}{\textit{e.g.}\xspace}
\newcommand{\ie}{\textit{i.e.}\xspace}
\newcommand{\paper}{article\xspace}

\newcommand{\Intel}{Intel$\textsuperscript{\textregistered}$\xspace}
\newcommand{\SKX}{Xeon$\textsuperscript{\textregistered}$\xspace Platinum 8160}
\newcommand{\KNL}{Xeon Phi$\textsuperscript{\textrm{TM}}$\xspace 7250}

\begin{document}

\title{High-Performance Code Generation though Fusion and Vectorization}

\author{Jason Sewall\\ \href{mailto:jason.sewall@intel.com}{jason.sewall@intel.com} \and Simon J. Pennycook \\ \href{mailto:john.pennycook@intel.com}{john.pennycook@intel.com}}
\date{Intel Corporation\\ \vspace{0.5cm} \today}
% \affiliation{%
%   \institution{Intel Corporation}
%   \streetaddress{2200 Mission College Boulevard}
%   \city{Santa Clara}
%   \state{CA}
%   \postcode{95054}
% }

\maketitle
\section*{Abstract}

We present a technique for automatically transforming kernel-based computations in disparate, nested loops into a fused, vectorized form that can reduce intermediate storage needs and lead to improved performance on contemporary hardware.

We introduce representations for the abstract relationships and data dependencies of kernels in loop nests and algorithms for manipulating them into more efficient form; we similarly introduce techniques for determining data access patterns for stencil-like array accesses and show how this can be used to elide storage and improve vectorization.

We discuss our prototype implementation of these ideas---named HFAV---and its use of a declarative, inference-based front-end to drive transformations, and we present results for some prominent codes in HPC.

%%% Local Variables:
%%% mode: latex
%%% TeX-master: "ms"
%%% End:

\section{Introduction}
\label{sec:introduction}

Hardware evolves, and languages can evolve with it, but codes remain static unless rewritten.  This \paper{} presents a method for automatically generating stencil-like codes---based on existing code and simple rules---that run well on modern hardware.
\subsection{Complex solutions, simple languages}
The longevity of many programming languages has been both a boon and a hindrance to computing.  Fortran, famously the first high-level programming language\todo{need a reference?}, has been in continuous use since 1957. C and C++ first appeared in 1972 and 1983, respectively\todo{need a reference?}.  Billions of lines of code have been written in these languages, providing a huge library of code for posterity, and their longevity means that developers have had many years in which to become familiar with the languages.

These languages were developed on and for hardware long deemed obsolete, in computing environments that would be nearly unrecognizable to most programmers today. New ideas about programming have come (and some have gone) during their lifetimes. All of this is to say that the sort of programs that run best on modern hardware, in modern computing environments, can be difficult or impossible to express in old languages.  To address this, older languages are revised and updated; Fortran has seen major revisions roughly every decade, as has C, and C++ seems to be on a 3- or 4-year cadence. \todo{need a reference?}

The modernization of these languages can help address many problems, but it largely cannot retroactively, automatically modernize older codes.  That mass of `legacy' code has inertia and---if the analogy may be stretched a little more---perhaps even a gravitational field; new codes related to a long-lived project may be developed in an older variant of a language to maintain compatibility.

Of course, writing a code from scratch in a modern language doesn't guarantee a straightforward, simple solution.  The complexity of modern hardware is such that complex software solutions are sometimes necessary to achieve the highest level of performance. Compilers have become extremely powerful and remain essential tools for developing efficient codes; even so, they are not always capable of transforming user code into the most efficient implementation for a given machine. This is where intermediate tools---optimization frameworks, libraries, domain-specific languages (DSLs) and source-to-source techniques like the one we present here---can be most useful.
\subsection{Contributions}
We present a method for generating efficient computations on regular grids based on user-specified code and rules. At its core, our technique aggressively fuses user-provided loops---potentially nested---to maximize locality and re-use of data; it is also able to reduce or eliminate storage of intermediate values, and it is capable of further transformations to enable higher-performance vectorization.

We have developed a prototype tool named High-performance Fusion And Vectorization (HFAV) that implements the techniques described in this \paper{} that accepts declarative input in the form of initial `axioms', terminal `goals', and a collection of production rules that describe individual kernels in the code and their data dependencies.  HFAV is then able to perform inference on this logical system to discover the dimensionality and dataflow of the program, to which the above techniques are applied to produce generated code that invokes the user-specified kernels. This is amenable to integration with an outer level of thread-level or SPMD-level parallelism, and our prototype has complete backends in C and C++ with preliminary support for Fortran.

This prototype has been open-sourced and released; see Section~\ref{sec:implementation} for details.
\todo[inline]{Tease results}
%
%%% Local Variables:
%%% mode: latex
%%% TeX-master: "ms"
%%% End:

\section{Related Work}
\label{sec:related-work}

\noindent Transforming user code into a form that is highly-tuned for a specific hardware architecture is an active area of research.  This is especially true in the domain of stencil codes, where operators and data access patterns are simple to define and characterize, but difficult to map to hardware such that they run efficiently (\ie accounting for node/thread/vector parallelism and all levels of the memory hierarchy).  As a result, the developers of stencil codes have a variety of actively developed DSLs and tools to choose from---such as STELLA~\cite{STELLA}, Pochoir~\cite{Pochoir}, PATUS~\cite{PATUS}, OPS~\cite{OPS}, PLuTo~\cite{PLUTO}, and YASK~\cite{YASK}---which are capable of automating parallelization, cache blocking, vectorization and changes in data layout.  Our technique employs optimizations that are orthogonal to those implemented in existing stencil frameworks, and assumes that something else (\ie the developer or another tool) is responsible for parallelization and data layout; we believe that the optimizations presented in this \paper are compatible with existing frameworks, but exploration of this is beyond the scope of this work.

The loop fusion and storage/bandwidth reduction optimizations from our technique are similar to some that can be found in other research, but they differ in their implementation.  Compiler passes that perform these optimizations automatically have been implemented~\cite{Ng-Contraction,Polly}, but their usage is limited to cases in which the compiler is able to recognize that the optimization is both valid and beneficial.  Both the OPS framework~\cite{OPS-Fusion} and the loop chain directives developed by Bertolacci et al.~\cite{Bertolacci-LoopChains} sidestep this problem by fusing loops based on user-provided data dependency information, but neither specifically targets the optimization or removal of intermediate storage locations.  The ``storage folding'' optimization in Halide~\cite{Halide} is most similar to that discussed here, with the key difference being our technique's specific consideration for efficient vectorization.

To the best of our knowledge, the use of an inference system to discover and satisfy data dependencies that is found in our prototype implementation is novel.  However, the use of inference systems in general is not novel in this space.  One example is the inference system used by PetaBricks~\cite{PetaBricks} to construct algorithms from device- and context-specific building blocks.  HFAV's inference system is currently much simpler that PetaBricks'---allowing for only one function to produce a given output---and would likely need to be extended if our implementation were to support a wider range of architectures.

%%% Local Variables:
%%% mode: latex
%%% TeX-master: "ms"
%%% End:

\section{Design}
\label{sec:design}

Some terminology we use in this \paper{} is informed by the declarative, rule-based front-end found in our prototype implementation. While this front-end is useful, the fusion and transformation techniques at the core of this method do not depend on it; we describe our prototype and its front-end, as well as other front-ends in progress, in
Section~\ref{sec:implementation}.

To help convey the key concepts of our technique, we employ various example codes and demonstrate their transformation process through the code generator.  The 5-point Laplace stencil is one such example; this is a common finite-difference discretization of the Laplace operator in two spatial dimensions. Listing~\ref{code:laplace5} shows the Laplace stencil used in a successive-over-relaxation (SOR) method.

\begin{figure}
\begin{lstlisting}
void laplace5(float n, float e, float s, float w,
              float c, float* out)
{
  *out = (1.0f-om) * c + om/4.0f*(n + e + s + w);
}

void apply_laplace5(int N, const float in[][N+2],
                    float out[][N+2])
{
  for (int j = 1; j < N+1; ++j)
    for (int i = 1; i < N+1; ++i)
      laplace5(in[j-1][i], in[j][i+1], in[j+1][i],
               in[j][i-1], in[j][i], &out[j][i]);
}
\end{lstlisting}
\caption{Simple code for the 5-point Laplace stencil}
\label{code:laplace5}
\end{figure}

We also use a code that performs one-dimensional flux differences on a system of equations that needs to normalize each computed flux vector. This example is greatly simplified from real-world hydrodynamics codes, but shows important data flow concepts.

\subsection{Fundamentals}
\label{sec:fundamentals}

Our technique operates on codes that apply homogeneous computations over \emph{iteration spaces}---multidimensional outer products of series of equally-spaced integers.  In imperative terms, we operate on codes that call kernels in successive nested `simple'\footnote{`Simple' here means that each loop advance a single integral counter by a fixed integral stride. The number of trips must not depend on other enclosing or enclosed loops or kernel computations.} loops. We reference the dimensions of this iteration space with familiar loop labels such as $i$, $j$, and $k$.

While the computations found in these nested loops need not be independent of one another---either within a loop nest or across them---our transformations require that the code be agnostic to the order in which a given kernel is applied; this freedom to re-order computations is critical in transforming the code. This implies that we do not allow kernels to have `side-effects' (\ie{} effects on global program state).

Kernels are described in terms of their data accesses/outputs against a canonical frame of reference. So, while the Laplace code shown above accesses a 2D data grid of size $N \times N$, it is convenient to talk about it in a translation-free sense as an operator using relative $\pm1$ offsets. This is in fact very much the way the invocation to the \texttt{laplace5} appears in Listing~\ref{code:laplace5}.

When performing fusion analysis, we impose a user-selected global loop ordering on the code: while kernels can be expressed in an order-free listing of iteration space depth (\eg{}, $(i, j)$ or $(j, i)$), ultimately they are arranged to match the global order to enable fusion.

Our transformation technique follows the following steps:
\begin{enumerate}
\item Form dataflow graph
\item Build iteration nests
\item Build iteration nest graph
\item Fuse iteration nest graph
\item Analyze variables\\
  (in-outs, reuse, contraction, vectorization)
\item Emit code
\end{enumerate}

These steps are described in detail below.

\subsection{Dataflow}
\label{sec:dataflow-dag}

\begin{figure}
    \centering
  \includegraphics[width=0.8\columnwidth]{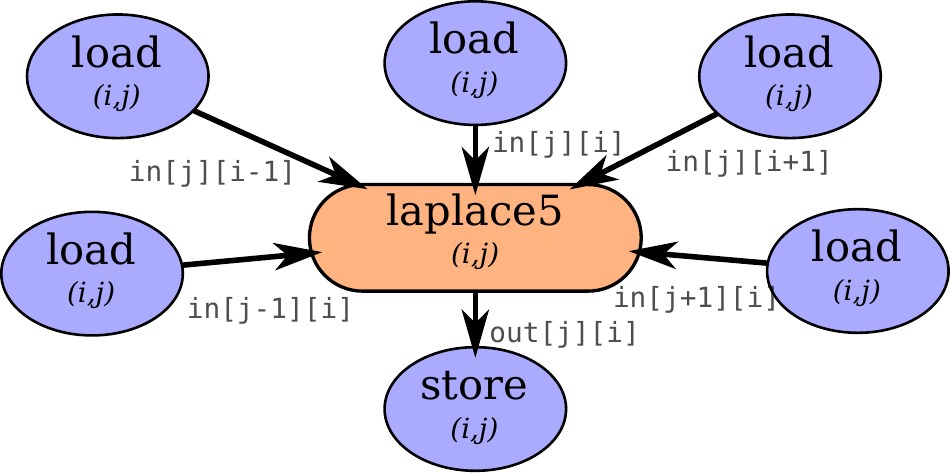}
  \caption{Dataflow DAG for the 5-point Laplace stencil.}
  \label{fig:laplace-dataflow}
\end{figure}

We begin with a dataflow graph provided by the front-end; this is a directed acyclic graph (DAG) with kernel callsites as vertices and intermediate variables forming the edges. Such a graph for the Laplace example is shown in Figure~\ref{fig:laplace-dataflow}. Note that we have added pseudo-kernels such as `load' and `store' to handle terminal references.

\begin{figure}
    \centering
  \includegraphics[width=0.8\columnwidth]{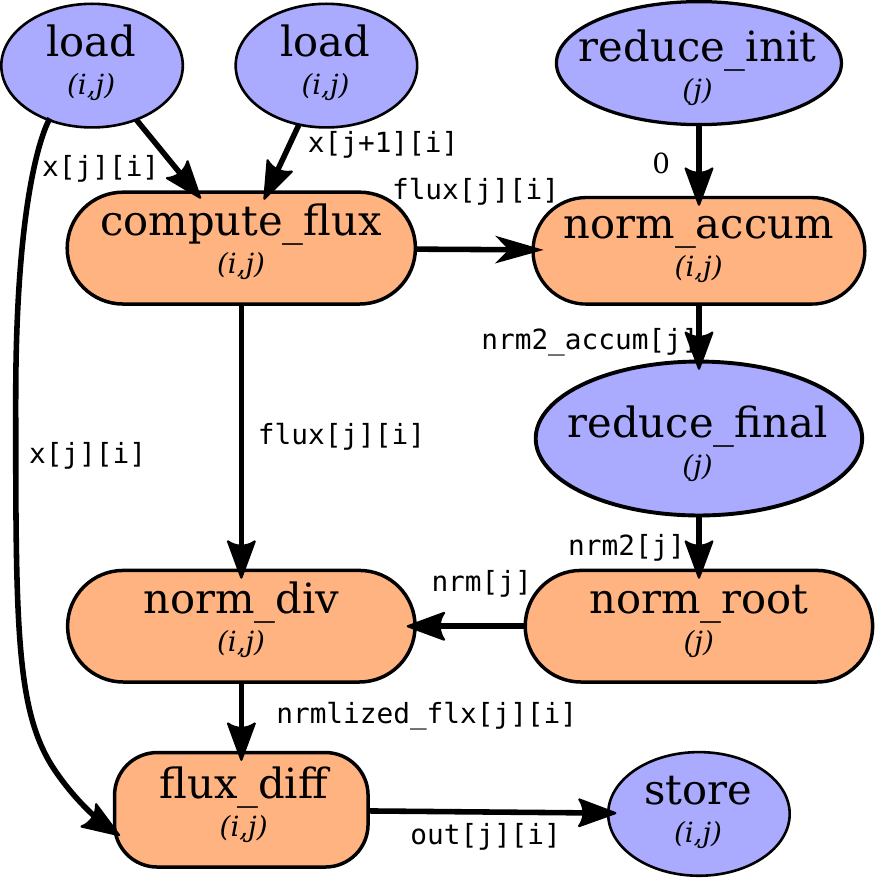}
  \caption{Dataflow DAG for the normalization example.}
  \label{fig:adv_norm-dataflow}
\end{figure}

A significant amount of information may be gleaned from this graph; in addition to informing the eventual code generation (through a topological traversal of this graph), we can determine the iteration space for each kernel  callsite by taking the union of all iteration spaces found on incident variables.

To enable fusion and other transformations, we will use this graph to build a data structure more suited to the task. First, we introduce a building block.

\subsubsection{Iteration nests}

An iteration \emph{space} describes all trips taken by a set of tightly nested loops, such as the loops found in Listing~\ref{code:laplace5} where there is no code between the two \texttt{for} statements. This is a useful construct, but it cannot describe some iteration patterns that are useful to manipulate.

A common pattern in performing iteration is to perform some work before a loop (the \emph{prologue}), the main loop body itself (the \emph{steady-state}), and to perform some work after the loop (the \emph{epilogue}). We refer to these steps as the \emph{phases} of the iteration nest. A single iteration nest has an associated identifier, range, and stride that is used to to generate code and to determine compatibility during fusion.

The prologue can be omitted if the kernel invocations and iteration nests therein are the same as those in the steady-state; the same applies to the epilogue; and when all phases are identical, the `perfect' iteration nest directly corresponds to an iteration space. Note that it is possible to permute the nesting of perfect iteration nest, which is equivalent to permuting the order of loops; this is useful for imposing a global loop nesting order on kernel callsites.

The general form of an iteration nest with distinct phases and recursive structure (\ie{} other iteration nests in one or more of those bodies) takes the form of a $[1,4)$-ary tree. It is sometimes useful to refer to the maximum depth of a given tree when performing fusion.

\subsubsection{Iteration nest DAG}

\begin{figure}
  \centering
  \includegraphics[width=0.8\columnwidth]{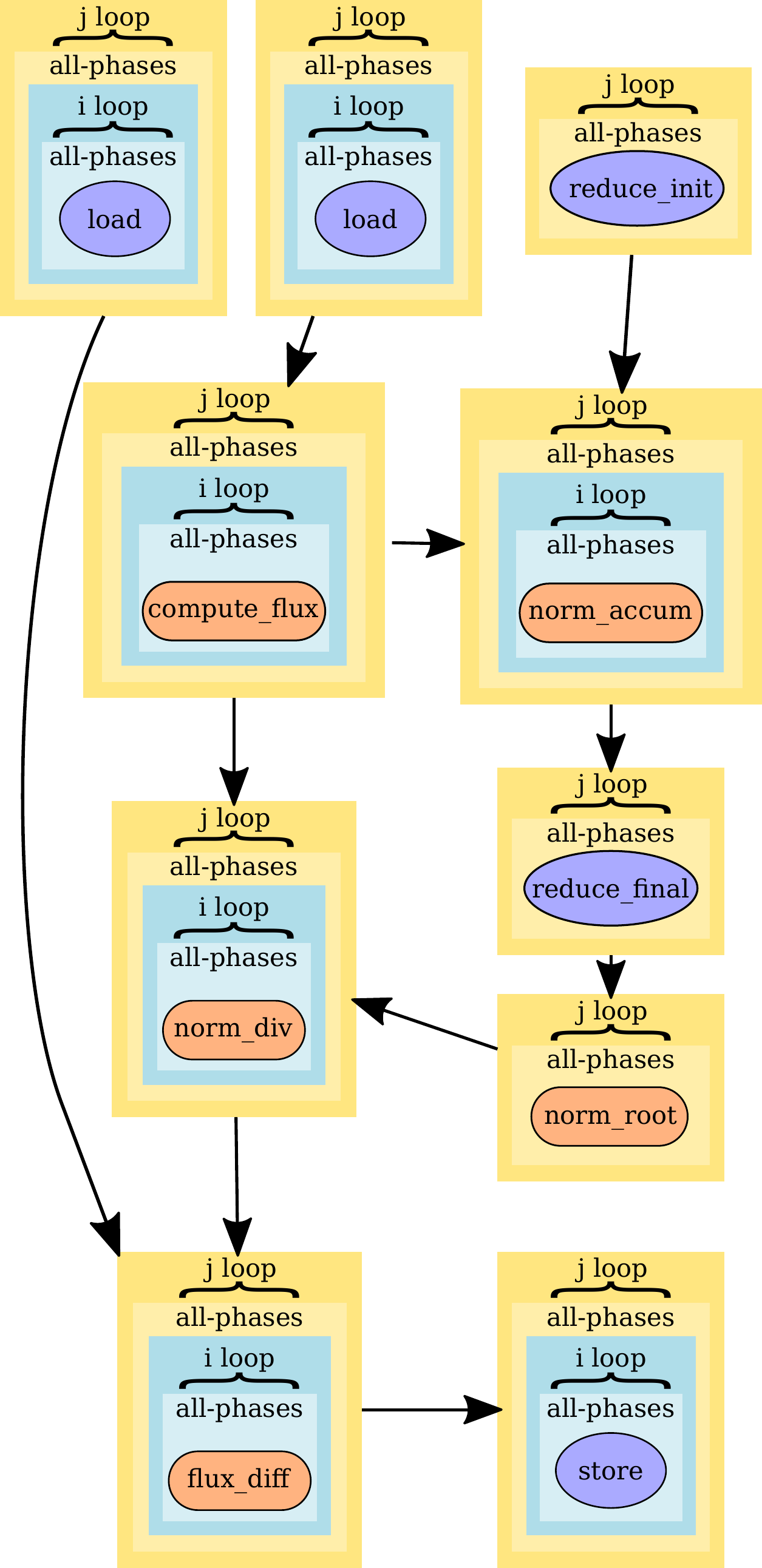}
  \caption{Initial iteration nest for the normalization example.}
  \label{fig:normalization-inest}
\end{figure}

The dataflow graph's structure can generate the iteration space for each kernel callsite (see Section~\ref{sec:dataflow-dag}); this can be used to initialize a perfect iteration nest that we will manipulate in later stages.  We form the initial iteration nest DAG by first aggregating certain kernel callsites (see `Grouping', below) by merging vertices, and then creating the aforementioned perfect iteration nests from those groups with callsites of the innermost nest steady-states. See Figure~\ref{fig:normalization-inest} for the initial iteration nest DAG from the normalization example.

\paragraph{Grouping} \label{sec:grouping}

While it is possible to construct the iteration nest DAG from the dataflow DAG directly (without merging any vertices) it makes sense to combine related kernel callsites. Specifically, we examine the kernel callsites and group them based on: (1) matching kernel names, and (2) matching parameter lists. A parameter list is said to match if the parameters are identical in value and position except for the spatial displacements. So \{\verb:load(cell[i][j]):, \verb:load(cell[i+1][j]):, \verb:load(cell[i-1][j]):, \verb:load(cell[i][j-1]):\},  \verb:load(cell[i][j+1]):\}---differing only in the displacements in \texttt{i} and \texttt{j}---are grouped together, while \{\verb:load(cell[i][j]):, \verb:load(aux[i+1][j]):\} would not be.

\subsection{Fusion}
\label{sec:fusion}

Once the iteration nest DAG is constructed, we can begin to fuse the iteration nests located at vertices. Fusion has two processes: (1) an outer level where certain pairs of vertices of the iteration nest DAG are examined to determine if they may be fused; and (2) an inner level where, if fusion is to occur, the two iteration nests at said vertices are fused.

\subsubsection{Fusing the iteration nest DAG}

The outer level of this process is a traversal of the iteration nest DAG that maintains a `fusing' vertex (which may be the fused aggregate of many of the vertices the original inest DAG); fusion is attempted across each incoming edge, which eventually results in the exhaustion of incoming edges or an unfusable \emph{split} (see below). Pseudocode for this procedure is given in Listing~\ref{code:inest-dag-fusion}.

\begin{figure}
  \begin{lstlisting}
procedure fuse_inest_dag(inest_dag)
  for vert in topo_sort_vertices(inest_dag)
    for iedge in incoming_edge_list(vert)
      fuse_result = fuse_inest(vert,
                               incoming_term(iedge))
      if fuse_result == unfusable
        for edge in sep_edges(incoming_term(iedge),
                              vert)
          split(edge)
      else
        vert = fuse_result
  \end{lstlisting}
  \caption{Algorithm for iteration nest DAG fusion}
  \label{code:inest-dag-fusion}
\end{figure}

The traversal must occur in a topological order to ensure that we do not introduce cycles when fusing vertices.  When an edge is identified as unfusable, we identify all vertices reachable from the candidate vertex and cut the iteration nest DAG along the separating edges between that subgraph and its complement.

\begin{figure*}
  \centering
  \includegraphics[width=0.8\textwidth]{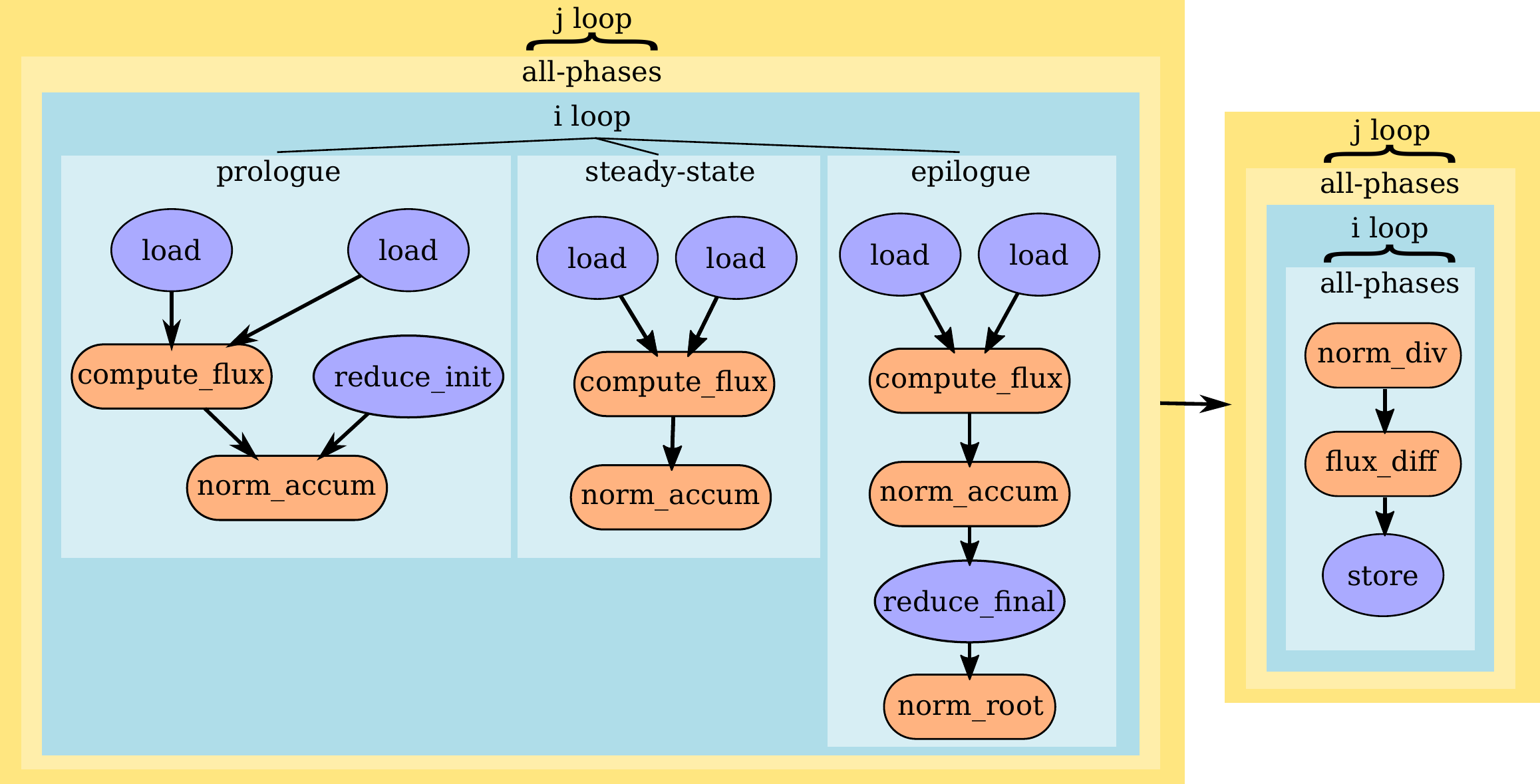}
  \caption{Fused iteration nest for the normalization example.}
  \label{fig:normalization-inest-fused}
\end{figure*}

The results of fusion applied to the iteration nest DAG from Figure~\ref{fig:normalization-inest} are shown in Figure~\ref{fig:normalization-inest-fused}. Here, the entire iteration nest DAG fused into a single iteration nest.

\subsubsection{Fusing iteration nests}

The \texttt{fuse\_inest()} procedure in Figure~\ref{code:inest-dag-fusion} handles the details of fusing iteration nests.

\paragraph{Rank ordering} First, recall (see Section~\ref{sec:fundamentals}) that we have established a global ranking of loop identifiers (\eg{}, a triple-nested loop might have identifiers ordered as $(k, j, i)$: $k$ is rank 2 and outermost, $j$ is rank 1, and $i$ is rank 0 and the innermost).  By construction, the identifiers within iteration nests obey this order, and each iteration nest $A$ has an $\mathtt{irank}(A)$ equal to the rank of its outermost identifier---\eg{}, following the previous example, a double-nested iteration nest $A$ over $k$ and $i$ would have $\mathtt{irank}(A)=2$, and a double-nested iteration nest $B$ over $j$ and $i$ would have $\mathtt{irank}(B)=1$.

\paragraph{Dataflow ordering} Second, recall that each iteration nest contains---in its leaves---nodes from the dataflow DAG.  For any two iteration nests, or phases within iteration nests, we can examine the relative ordering of the dataflow DAG subgraphs induced by these iteration nests. Given two such subgraphs $\mathcal{R}$ and $\mathcal{S}$ of the dataflow DAG $\mathcal{D}$, we can ask if $\left. \left(\mathcal{R} \le \mathcal{S}\right) \right| \mathcal{D}$, which asks if each node of $\mathcal{R}$ can be topographically ordered before each node of $\mathcal{S}$ in $\mathcal{D}$.  For two such subgraphs, there are four possiblities:

\begin{enumerate}
  \item $\left. \left(\mathcal{R} \le \mathcal{S}\right) \right| \mathcal{D}$ = true and $\left. \left(\mathcal{S} \le \mathcal{R}\right) \right| \mathcal{D}$ = true: $\mathcal{R}$ and $\mathcal{S}$ are indepdendent of one another in $\mathcal{D}$ and maybe be relatively ordered in any fashion.
  \item $\left. \left(\mathcal{R} \le \mathcal{S}\right) \right| \mathcal{D}$ = true and $\left. \left(\mathcal{S} \le \mathcal{R}\right) \right| \mathcal{D}$ = false: $\mathcal{R}$ must be ordered before $\mathcal{S}$.
  \item $\left. \left(\mathcal{R} \le \mathcal{S}\right) \right| \mathcal{D}$ = false and $\left. \left(\mathcal{S} \le \mathcal{R}\right) \right| \mathcal{D}$ = true: $\mathcal{S}$ must be ordered before $\mathcal{R}$.
  \item $\left. \left(\mathcal{R} \le \mathcal{S}\right) \right| \mathcal{D}$ = false and $\left. \left(\mathcal{S} \le \mathcal{R}\right) \right| \mathcal{D}$ = false: $\mathcal{R}$ and $\mathcal{S}$ have a cycle between them and cannot be ordered; even though $\mathcal{D}$ is acyclic by definition, subgraphs of it need not maintain that property.
\end{enumerate}

Fusion operates on pairs of iteration nests (we shall say $A$ and $B$) using these concepts of rank ordering and dataflow ordering, and is applied recursively---a natural consequence of the recursive nature of iteration nests. First, we compare the rank order: $\mathtt{irank}(A)$ and $\mathtt{irank}(B)$; this determines how kernels in the phases of $A$ and $B$ may be combined. When the ranks match (\ie{}, the outer-most identifiers are the same), we recursively fuse each phase of $A$ and $B$ directly (\ie{} we fuse the prologue of $A$ to the prologue of $B$, we fuse the steady-state of $A$ to the steady-state of $B$, and we fuse the epilogue of $A$ to the epilogue of $B$), subject to the ability to find a compatible dataflow order, as described above.

When the ranks of $A$ and $B$ differ, we fuse the lower-ranked iteration nest's phases into the higher-ranked iteration nests's prologue or epilogue, depending on the dataflow ordering. In any case, if we are unable to establish an order between iteration nests, we have identified the need to split. Pseudocode for iteration nest fusion is given in Listing~\ref{code:inest-fusion}.

\begin{figure}
  \begin{lstlisting}
procedure fuse_inest(A, B)
  // irank() gives the rank of the outermost loop
  // Here, it is used to (try to) determine a nesting
  // order
  diff = irank(A) - irank(B)
  if diff == 0
    // dataflow_le() returns 'true' if the second
    //  argument can be toplogocially ordered after
    //  the first
    // {prologue,epilogue}_only() return the kernel
    //  callsites of
    //  the all children in the {prologue,epilogue}
    //  minus those in the steady_state()
    if dataflow_le(prlg_only(A), steady(B)) and
       dataflow_le(prlg_only(B), steady(A)) and
       dataflow_le(steady(A), eplg_only(B)) and
       dataflow_le(steady(B), eplg_only(A))
       // inest(iter_ident, [prlg, steady, eplg])
       // creates a new iteration nest with the given
       // identifier and phases
       return inest(iident(A),
                    [fuse_inest(prlg(A), prlg(B)),
                     fuse_inest(steady(A), steady(B)),
                     fuse_inest(eplg(A), eplg(B))]
    else return unfusable
  else if diff < 0:
    swap(A, B)
  before? = dataflow_le(all_phases(A), steady(B))
  after? = dataflow_le(eplg(B), all_phases(A))
  if before?
    // if after? is true, fusion order is ambiguous
    return inest(iident(A),
                 [fuse_inest(all_phases(A), prlg(B)),
                  steady(B),
                  eplg(B)]
  else if after?
    return inest(iident(A),
                 [prlg(B),
                  steady(B),
                  fuse_inest(all_phases(A), eplg(B))]
  else return unfusable
  \end{lstlisting}
  \caption{Algorithm for recursively fusing two iteration nests. Invoked from \texttt{fuse\_inest\_dag} in Figure~\ref{code:inest-dag-fusion}.}
  \label{code:inest-fusion}
\end{figure}

\subsection{Reductions \& broadcasts \& splits}

Most real codes are relatively simple to fuse: for example, a 7-point Laplace stencil might consume three-dimensional data from different offsets and produce three-dimensional data. Flux differencing codes consume adjacent pairs of three-dimensional data to compute fluxes, then those fluxes are used in pairs or triples for limiting and integration into a three-dimensional output. When the inputs and outputs to kernels are all of the same dimension, their loops don't require special handling.

However, there are real use cases where a kernel might consume one-dimensional data to produce three-dimensional data, or vice versa. These require special consideration. We refer to kernels that transform lower-dimensional data to higher-dimensional data as \emph{broadcasts} and to those that fransform higher-dimensional data to lower-dimensional data as \emph{reductions}.

Broadcasts can be handled by fusing the producer of the lower-dimensional data into the prologue of one of the consumers' iteration nests.

Reductions require more consideration because they involve many writes to the same data; because of our iteration-order-agnostic assumptions, valid reduction kernels must implement associative operators. They are typically found in a three-step pattern spread across three kernels: first, an initialization kernel sets the accumulation data to the identity of the associative operator, then the associative kernel accumulates over that data, and last a finalization kernel is run to finish the result. For example, in computation of the mean across a one-dimensional array, you would first set an accumulator to zero, then iterate over the array, adding entries to the accumulator and counting steps, and finally you would divide the accumulator by the number of steps taken. This generalizes to averages across columns or rows of two-dimensional data and so on.

These triples fit nicely into the phase scheme outlined earlier: initialization forms the prologue of the fused iteration nest, the associative kernel forms the steady-state, and the finalization forms the epilogue.

\paragraph{Splits}

Naturally, it is desirable to fuse iteration nests containing broadcasts and reductions as much as possible to promote locality of reference and reuse. For most cases, iteration nests containing reductions and broadcasts can be handled as any other; the exception is when a broadcast consumes the result of a reduction. Such regions have dataflow from higher dimensions to a lower one (the reduction), and then back up to a higher dimension (the broadcast).  We refer to this as \emph{concave dataflow} and the need to handle the lower-dimensional data property prevents the two higher-dimensional iteration nests (which need not be of the same depth) from being fused. The result is a \emph{split}.

The identification of a split not only prevents the two iteration nests that gave rise to the split from being fused, but requires that we cut the dataflow graph to separate the two subgraphs---one upstream, one downstream---induced by the  identified split. Those kernels downstream of the split depend on the results therein, and therefore we must separate the graph completely.

\subsection{Variables and storage}

The variable references from the dataflow DAG can be used to determine the storage necessary to execute each kernel correctly. The size of each iteration nest, combined with the total offsets used to reference each variable, can be used to determine the size of each dimension of access.\footnote{This is similar to the Minkowski Sum found in geometry.}

One advantage of fusion is that it allows for storage to be reduced because of improved locality of reference; we can compute the precise number of times a value is used and schedule loops to minimize intermediate storage.

\paragraph{Enclosing}

In most cases, variables end up having limited scope due to their producers and consumers being fused into the same iteration nest. However, when fusion is not possible (because of splits, see above) data must be kept live across distinct iteration nests.

To effect this, we analyze dataflow in the fused iteration nest to find the narrowest region of liveness to use for each variable; we refer to this as the \emph{enclosing} region for the variable. Most such regions are trivial, being internal to a single iteration nest, but others span multiple iteration nests.

\paragraph{Grouping}

The first step in reducing storage of intermediate values is to identify reuse patterns. This can be done by examining all variable references used as inputs and aggregating those going to the same identifier but with distinct references. For example, a 5-point Laplace stencil access 5 offsets in a single array: $(i,j), (i+1,j), (j+1, i), (i-1,j), \textrm{and}\,(i,j-1)$.

With references aggregated, we are then able to use the iteration nest's iterator order to find reuse patterns for each reference. In the Laplace example, if we proceed with iterating in $i$, $j$ order in linear fashion, we know that when we access $(i,j+1)$, it is the first time we have visited that value as input. The next time we access that variable will be at $(i+1,j)$, and it will be accessed next by $(i,j), (i-1,j)$, and finally $(i,j-1)$ before it is not needed again.

\begin{figure}
  \centering
  \def\svgwidth{0.5\columnwidth}
  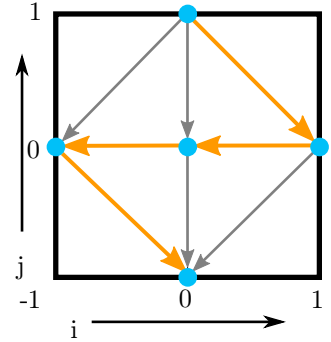
  \caption{Reuse diagram for the 5-point Laplace stencil. Arrows show reuse patterns for an $(j,i)$ progression; the orange arrows show the Hamiltonian path of reuse.}
  \label{fig:laplace5-reuse}
\end{figure}

Figure~\ref{fig:laplace5-reuse} shows this reuse pattern.  This pattern can be produced procedurally in graph form for each identifier $v$ by a three-step process: (1) creating vertices for each reference to $v$; (2) adding an edge from $a$ to $b$ where $a$ appears before $b$ in the order induced by the iteration ordering; and finally, (3) computing the longest path in the graph, which will produce a Hamiltonian path covering all nodes in the order that they will be reused.

\paragraph{Contraction}
\label{sec:contraction}
With reuse patterns computed, it is possible to contract the storage needed for each input variable by determining, for each identifier in an iteration nest, the distance in iterations between the first and last reference for each variable.

For a 1D, 3-point Laplace example that references the array $u$ at $i+1,i,i-1$, grouping analysis shows us that the first time we see a value in the steady-state is at $i+1$. It is reused by $i$ and then $i-1$, and then isn't needed again.

\begin{figure}
  \centering
  \subcaptionbox{Rotation scheme for data in the innermost dimension.\label{fig:inner-rotation}}{\includegraphics[width=0.5\columnwidth]{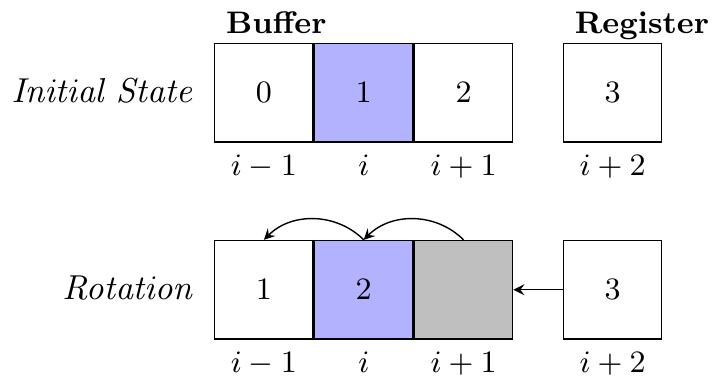}}
  \subcaptionbox{Rotation scheme for data in a dimension that is not the innermost.\label{fig:outer-rotation}}{\includegraphics[width=0.7\columnwidth]{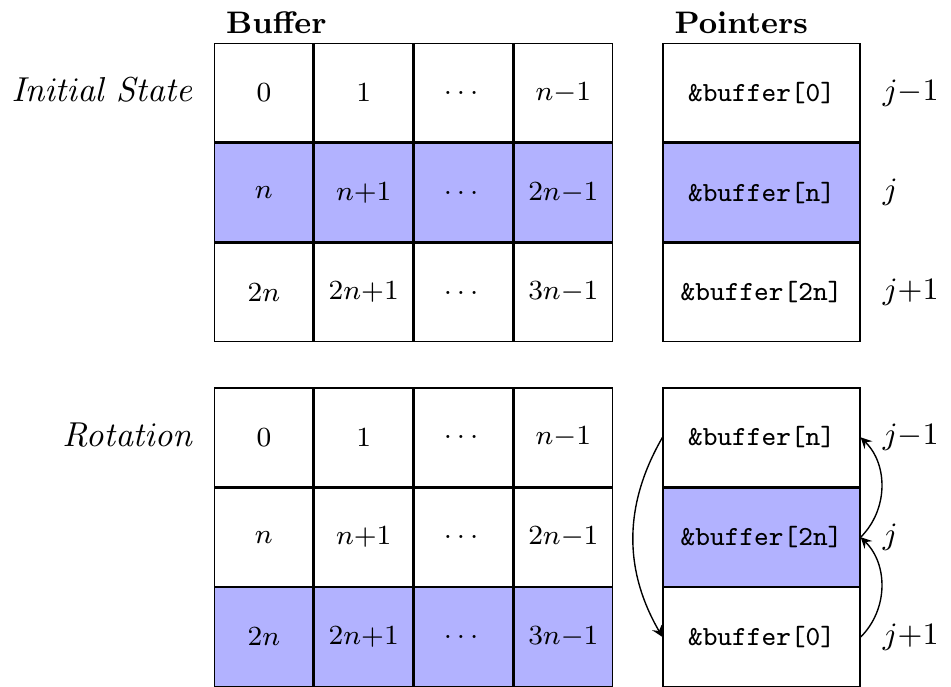}}

  \subcaptionbox{Vectorized rotation scheme for data in the innermost dimension, for a vector length of 4.\label{fig:vector-rotation}}{\includegraphics[width=0.8\columnwidth]{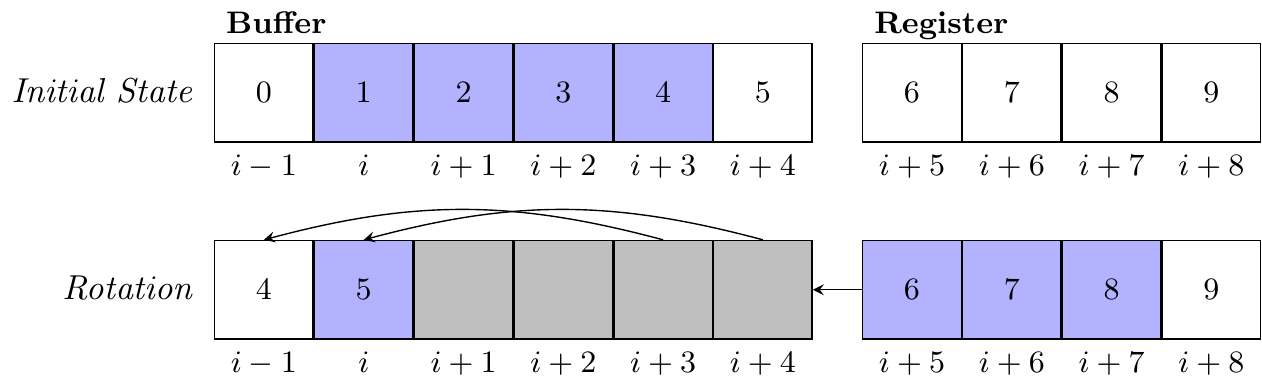}}
  \caption{Rotation schemes for storage contraction.}
\end{figure}

Na\"{\i}ve analysis would suggest that we need a buffer for $u$ that covers the entire iteration space for the Laplace operator, plus 2 for the $i-1$ and $i+1$. Distance computations from the contraction analysis show that storage for only 3 values is necessary, and that rotation of the values or use of a circular buffer can manage the updates to storage. See Figure~\ref{fig:inner-rotation} for a diagram of this pattern.

The same principle can be applied to multidimensional references. For the 5-point Laplace stencil in 2D, the reuse distance for the inputs to the stencil on a variable $u$ given iteration order $j,i$ span $(i,j+1)$ to $(i,j-1)$---essentially 2 times the storage necessary for a single $j$-row.  We can produce a circular buffer or rotation scheme to cover this, although it is generally most practical to simply allocate 3 times the storage needed for a single row and to quickly rotate those rows through pointer manipulation. Figure~\ref{fig:outer-rotation} shows this `outer' rotation scheme.

\paragraph{In/out chaining}

In most cases, the dataflow graph contains terminal data references that refer to storage external to the analysis and transformation process described here; these are effectively the inputs and outputs to the transformed code. It is possible for inputs to alias outputs (for example, in an in-place stencil update)---the user must report this as part of the transformation input, since our default assumption is that variables do not alias one another.

To preserve correct behavior in the presence of aliasing, we use the dataflow graph to chain from each terminal input across \emph{kernel} inputs and outputs to find which terminal outputs depend on which terminal inputs. Then, if there is aliasing in interdependent terminals, we determine which input offsets must be copied to temporaries before output can be written back safely.

\paragraph{Vectorization}
\label{sec:vectorization}
Vectorization is a powerful tool for improving performance on modern processors, and the assumptions that we impose upon inputs to our transformation technique (order-independent execution, no aliasing, no side effects) generally are very compatible with vectorization.

However, one challenge is that while vectorization is \emph{possible} for these codes, it is often not helpful when performance is bound by memory bandwidth; this is typical of codes with disparate loops with multiple streams and long reuse distances---precisely the kind of code that our technique seeks to transform.

Fusing loops, reducing storage footprint, and reducing reuse distance can improve the arithmetic intensity of code to the point where vectorization is beneficial, and generally doesn't impede the process of vectorization.

One hiccup here is that contracting storage (see above) into circular buffers results in data access patterns incompatible with vectorization; in this case, the circular buffers are expanded by a target vector length and vector code to perform an in-place rotation of the buffer can be emitted. See Figure~\ref{fig:vector-rotation} for a visual depiction of how circular buffers expand.

\subsection{Code Generation}

With the fused iteration nest DAG and full variable analysis completed, code generation can begin.

We visit the iteration nest DAG in topological order, emitting enclosing regions and their variable definitions as we encounter them. We emit a loop body for each iteration nest phase and recursively traverse children in an in-order fashion.  Any kernel invocations in a phase are emitted according to the topological order found in the original dataflow graph, with appropriate variable references substituted.

When leaving iteration nests, the loops are closed and any necessary variable rotations are emitted.

%%% Local Variables:
%%% mode: latex
%%% TeX-master: "ms"
%%% End:

\section{Implementation}
\label{sec:implementation}

Our prototype implementation of the methods described in this paper---named HFAV---is a research-quality tool that nonetheless is able to produce compelling and even surprising results. HFAV is written in Python for quick development and access to external libraries, but could easily be imagined in a lower-level, more performant language in a production environment.

We use a custom YAML format for handling user input. The front-end of our tool may appear to accept C-like input, and that may be a convenience for users coming from C, but it is foremost simply a way of declaring types, functions, and dependencies, and the mechanics of the way code is handled in the front-end are purely based on substitution. HFAV only needs to know the positions of arguments and the function name to emit compilable code.

The code emitted by HFAV can be included directly into programs or treated as stand-alone source files. The backend languages supported by our tool are C++ and C99, with some work on a Fortran 90 backend begun.

\begin{figure}
\begin{lstlisting}
kernels:
 laplace:
   declaration: laplace5(float n, float e, float s,
                         float w, float c, float &o);
     inputs: |
       n : q?[j?-1][i?]
       e : q?[j?][i?+1]
       s : q?[j?+1][i?]
       w : q?[j?][i?-1]
       c : q?[j?][i?]
     outputs: |
       o : laplace(q?[j?][i?])
globals:
  inputs: |
    float g_cell[j?][i?] => cell[j?][i?]
  outputs: |
    laplace(cell[j][i]) => float g_cell[j][i]
\end{lstlisting}
\caption{YAML inference system for the 5-point Laplace stencil}
\label{code:laplace5-yaml}
\end{figure}

\subsection{Inference}

HFAV analysis begins with a dataflow graph we refer to as the `inference DAG', or IDAG. This is a directed graph with individual data accesses---\emph{concrete terms}---as vertices and function calls and \emph{rule applications} or \emph{RAPs}---as edges.  Input terms form the roots of the IDAG, and output terms form the leaves.

\paragraph{RAP duals}

It is possible to define a dual of the IDAG that has RAPs as vertices and the terms interchanged between them as edges; we refer to this as the \emph{RAP dual}. This is in fact the same graph referred to as the `dataflow DAG' in Figure~\ref{fig:laplace-dataflow}.

\paragraph{Emitting code}

The availability of auto-vectorizing compilers and directives (like the \verb:omp simd: support found in OpenMP 4.0~\cite{OpenMP40}) means that our transformation can emit scalar loops that the compiler is able to vectorize.  For the cases where some special rotations are requires (see Secs.\ref{sec:contraction} and \ref{sec:vectorization}), we provide snippets of the appropriate language with intrinsic sequences for various instruction sets.

\paragraph{Debugging output}

One advantage of a Python implementation is that it is possible to easily emit various graphical outputs showing reuse patterns and dataflow; HFAV is capable of displaying these graphs at the users' request and is the basis for many of the figures in this \paper.

\paragraph{Availability}

We have released this prototype as an open-source project under the Apache License 2.0: \textbf{\url{https://github.com/intel/HFAV}}.

%%% Local Variables:
%%% mode: latex
%%% TeX-master: "ms"
%%% End:

\section{Performance}
\label{sec:performance}

To demonstrate the utility of HFAV in improving performance for bandwidth-bound codes, we present performance results and analysis for three different applications: the simple normalization example discussed throughout this \paper{}; a diffusion stencil; and a shock hydrodynamics benchmark.

\subsection{Experimental Setup}

\begin{table}
 \centering
 \scriptsize{}
 \begin{tabular}{r|p{2cm}|p{2cm}}
  & \textbf{SKX} & \textbf{KNL} \\
  \hline
  $\mathrm{Sockets}\times\mathrm{Cores}\times\mathrm{Threads}$ & $2\times24\times2$ & $ 1\times68\times4$ \\
  Ref. Clock (GHz) & 2.1 & 1.4 \\
  L1 / L2 /L3 Cache (KB) & 32 / 1024 / 32768  & 32 / 512 / - \\
  Peak DP GFLOP/s & 3225 & 3046 \\
  STREAM Bandwidth (GB/s) & 200 & 90 (DDR) \newline 490 (MCDRAM) \\
  \hline
 \end{tabular}
\caption{System configuration. L1 and L2 sizes are per-core; L3 is per-socket.}\label{tbl:config}
\end{table}

The system configuration used for all of our experiments is given in Table~\ref{tbl:config}.

We use the \Intel{} Compiler version 18.0, using the \texttt{-O3 -xHost} flags to request optimizations specific to the host on which the code is compiled.  We use all cores available on SKX and 64 cores on KNL (leaving 4 cores free for the operating system), launching two threads per core on both systems.  Thread affinities are set with \texttt{KMP\_AFFINITY=compact,granularity=fine} (to control thread placement) and \texttt{KMP\_HW\_SUBSET=\$\{cores\},2t} (which limits thread placement to specific subsets of logical cores).

The KNL is configured in `Quad/Flat' mode: its DDR and MCDRAM memories are exposed to software as two separate NUMA nodes, and the user must explicitly request allocations backed by MCDRAM (\eg{} through use of \texttt{libnuma} and its \texttt{numactl} utility).  We present results using MCDRAM only, thus highlighting that HFAV's data locality and bandwidth usage optimizations remain relevant even in the presence of the high-bandwidth memories available on some architectures.

% To highlight the impact of HFAV's optimizations on data locality and bandwidth usage, we present two sets of KNL results: one using DDR only (launched with \texttt{numactl -m 0}); and one using MCDRAM only (launched with \texttt{numactl -m 1}).

\subsection{Normalization Example}

After fusion (see Figure~\ref{fig:normalization-inest-fused}), the normalization example requires two loop nests: one containing the flux computation, norm accumulation and norm root; and another containing the final divisions and normalization operation.  The split between these two nests (required because of the reduction) prevents HFAV from performing array contraction optimizations---the data consumed by the second nest is produced by the first.

\begin{figure}
  \subcaptionbox{Throughput on 2 sockets of \Intel{} \SKX{}.}{  \centering\includegraphics{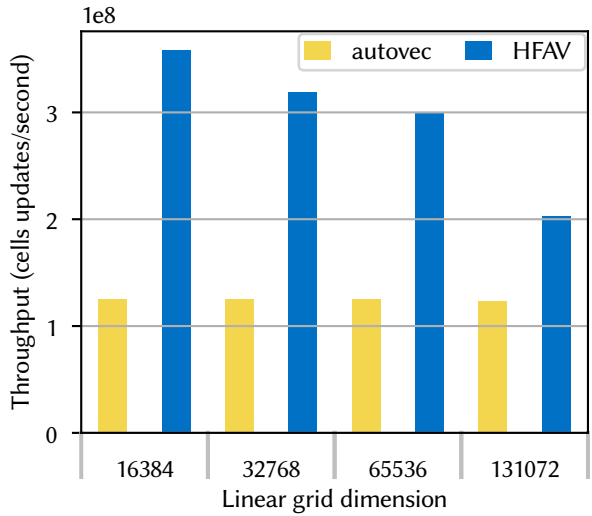}\label{fig:normalize-skx-results}}
  \subcaptionbox{Throughput on the \Intel{} \KNL{}.}{  \centering\includegraphics{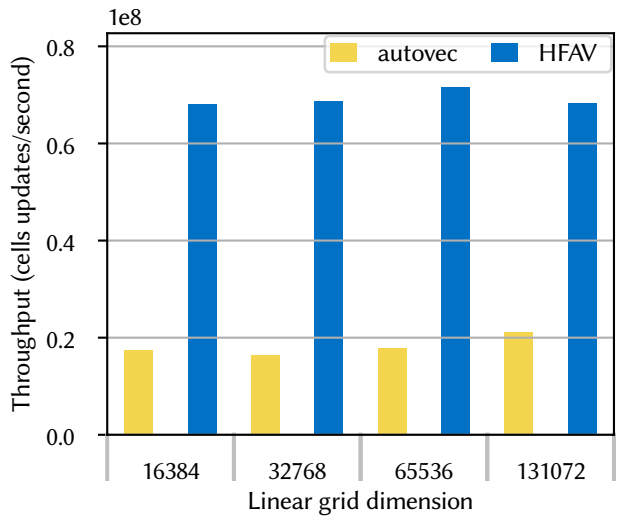}\label{fig:normalize-knl-results}}
  \caption{Performance of different implementations of the normalization example code for various platforms. `autovec' is the unmodified code (subject to whatever auto-vectorization optimizations the compiler performs), while `HFAV' is the output of our prototype generator.}\label{graph:normalize-results}
\end{figure}

The performance improvement arising from HFAV in this instance is therefore a direct result of loop fusion, which reduces the number of times that the entire $(j,i)$ space is visited from five to two.  As shown by the graph in Figure~\ref{graph:normalize-results}, this reduction in bandwidth requirements is reflected by the performance improvement that we see for large problems that do not fit in cache.

\begin{figure}
  \centering
  \includegraphics{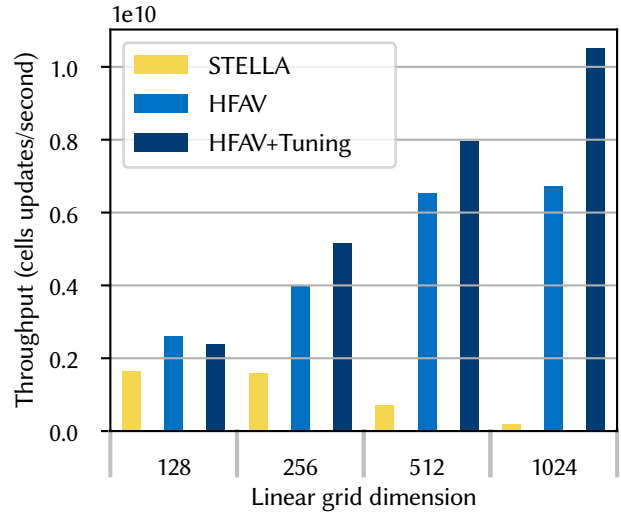}
  \caption{Performance of different implementations of the COSMO micro-kernels for \Intel{} \KNL{}. `STELLA' is the code from Gysi et al., `HFAV' is the (otherwise unmodified) output of our prototype generator, and `HFAV + Tuning' is the result of manual improvements to the output of `HFAV'.}\label{graph:COSMO}
\end{figure}

\subsection{COSMO Micro-Kernels}

In~\cite{STELLA}, Gysi et al.\ use a two-dimensional fourth-order diffusion stencil---a proxy for operations arising in the dynamical core of the COSMO atmospheric model~\cite{COSMO}---to demonstrate the functionality of the STELLA embedded DSL.\@ The stencil is applied over three-dimensional data (with no dependencies in one of the dimensions) and consists of four kernels performing very few floating-point operations per cell.

The four kernels are as follows:
\begin{itemize}
 \item \texttt{ulapstage}: 5-point Laplace operator, computed using the $u$ value for a cell and its neighbors in $i$ and $j$
 \item \texttt{flux\_x}: flux computation in $i$, using the $u$ values and the results of \texttt{ulapstage} for two neighboring cells in $i$
 \item \texttt{flux\_y}: flux computation in $j$ (as above, but in $j$)
 \item \texttt{ustage}: integration, using the $u$ value for a cell and the four neighboring fluxes in $i$ and $j$
\end{itemize}

The graph in Figure~\ref{graph:COSMO} shows performance results for several variants of the COSMO micro-kernels executing on KNL.  An optimized `STELLA' version provided by Gysi et al.\ fuses the final three kernels, with the fluxes computed redundantly for each cell.  The `HFAV' version merges all four kernels, using rolling buffers of sizes 2 and 3 for the fluxes and Laplacians respectively---in addition to reducing bandwidth requirements, memory footprint is reduced from $O(5N_{k}N_{j}N_{i})$ to $O(2N_{k}N_{j}N_{i} + 5N_i + 2)$.  Introducing an additional rolling buffer for the input values would permit the entire operation of this benchmark to be performed in-place, further reducing the memory footprint by $\approx2\times$; however, we omit this optimization from our study since we do not know whether it can be safely applied to COSMO.

As before, the performance impact of HFAV is largest for larger problem sizes, where intermediate results do not fit in cache.  In order to remain representative of COSMO, all problem sizes considered here are quite small---relative to the degree of parallelism available in modern machines---and therefore overheads introduced by core-to-core communication (\ie{} synchronization) and the software pipeline are exposed, limiting performance.  Our prototype assumes that the amount of work per loop iteration and the number of pipeline stages are both sufficiently large to amortize the costs of priming the software pipeline and accessing intermediate buffers.  This is not true for the COSMO micro-kernels when run on small problems, and we were able to improve upon our results by manually tuning the generated code ('HFAV + Tuning')---specifically, by using intrinsics to force alignment and instruction masking to fold the prologue/epilogue into the steady-state.  The manual nature of these optimizations reflects the immaturity of our prototype, and its current reliance on the compiler to auto-vectorize generated loops efficiently.

\subsection{Hydro2D}

Hydro2D is a two-dimensional shock-hydrodynamics benchmark developed by CEA~\cite{Hydro2D}.  The rolling optimizations performed by HFAV were first manually developed and tested in the context of Hydro2D, and we refer the reader to our previous work~\cite{Hydro2D-Pearls} for more details of that effort.

Hydro2D implements nine kernels:
\begin{itemize}
 \item \texttt{make\_boundary}: Set the boundary conditions.
 \item \texttt{constoprim}: Convert unknowns from `conservation form' to `primitive form'.
 \item \texttt{equation\_of\_state}: Complete the system of primitive equations.
 \item \texttt{slope}: Approximate derivatives of the averaged solution.
 \item \texttt{trace}: Limit slopes appropriately.
 \item \texttt{qleftright}: Split components according to cell boundaries.
 \item \texttt{riemann}: Solve Riemann problem at each boundary.
 \item \texttt{cmpflx}: Compute conservative fluxes based on Riemann solution.
 \item \texttt{update\_cons\_vars}: Integrate conservative variables in time.
\end{itemize}

The operator is dimensionally split (\ie{} it is applied in one dimension at a time), and each of its kernels therefore have dependencies in only one dimension.  Since the dimension in which a dependency occurs is different for each pass, HFAV effectively requires the user to specify the dependency information twice.

HFAV is able to fuse all nine of Hydro2D's kernels into a single loop nest, removing a significant number of intermediate buffers via array contraction in the process.  Storage for all but the four primitive variables ($\rho$, $\rho u$, $\rho v$ and $E$) can be replaced by rolling buffers with a maximum of 5 stages: memory footprint is thereby reduced from $O(31N_{j}N_{i})$ to $O(4N_{j}N_{i} + 112)$ in both the $i$ and $j$ passes.

\begin{figure}
  \subcaptionbox{Throughput on 2 sockets of \Intel{} \SKX{}.}{  \centering\includegraphics{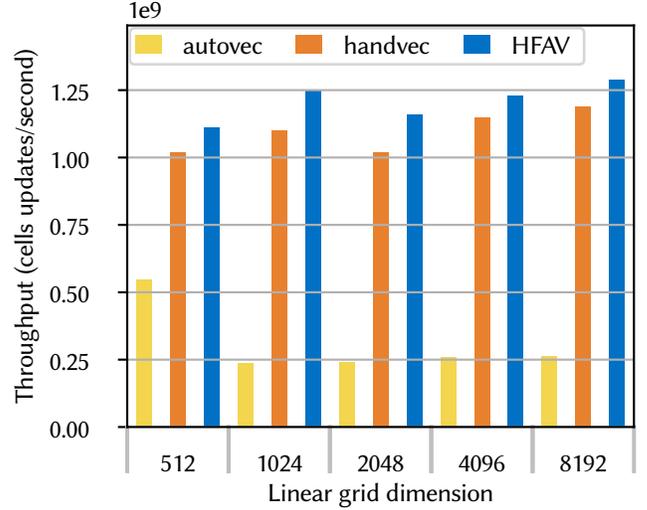}\label{fig:hydro2d-skx-results}}
  \subcaptionbox{Throughput on the \Intel{} \KNL{}.}{  \centering\includegraphics{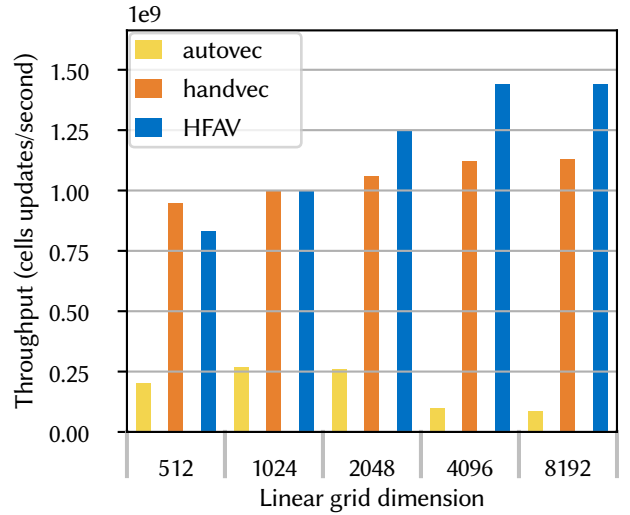}\label{fig:hydro2d-knl-results}}
  \caption{Performance of different implementations of the Hydro2D code for various platforms. `autovec' is the unmodified code (subject to whatever auto-vectorization optimizations the compiler performs), `handvec' is a manual vectorization using intrinsics, and `HFAV' is the output of our prototype generator.}\label{graph:Hydro2D}
\end{figure}

The graphs in Figure~\ref{graph:Hydro2D} show that HFAV is able to achieve performance similar to those of our manual efforts (from~\cite{Hydro2D-Pearls}); for problems where a significant number of kernels can be fused, the instruction overheads that effected the COSMO micro-kernels can be amortized.  Not reflected in the graph is the shift in architectural bottleneck after fusion; fusing all of the kernels and contracting the vast majority of the arrays moves Hydro2D from being bandwidth-bound to compute-bound---indeed, our generated code has almost identical performance on KNL whether using DDR or MCDRAM memory (results not shown).

%%% Local Variables:
%%% mode: latex
%%% TeX-master: "ms"
%%% End:

\section{Summary}
\label{sec:summary}

We have presented a technique for transforming input code consisting of kernels and loops into fused, vectorized kernels that can have more compact footprints and better data reuse properties. Our technique integrates with outer-level parallelization schemes such as OpenMP or MPI, and we have shown sizable speedups for important proxy codes on state-of-the art hardware.

We hope our technique lays the groundwork for expanded and more powerful code generation/transformation tools.

\paragraph{Limitations \& Future Work}

Some of the assumptions underlying our technique can be restrictive, and we would like to see future efforts address these restrictions. In particular, we would like to see direct support for automatic domain decomposition and threading, treatment of intermixed loop orderings, and the ability to fuse codes on unstructured grids.

We believe that the data dependency information required by HFAV could be expressed via directives and integrated into existing codes---similar to the approach in~\cite{Bertolacci-LoopChains}---and are interested in exploring this approach.

%%% Local Variables:
%%% mode: latex
%%% TeX-master: "ms"
%%% End:

\bibliographystyle{plain}
\bibliography{ms}

\begin{thebibliography}{10}

\bibitem{PetaBricks}
Jason Ansel, Cy~Chan, Yee~Lok Wong, Marek Olszewski, Qin Zhao, Alan Edelman,
  and Saman Amarasinghe.
\newblock {PetaBricks: A Language and Compiler for Algorithmic Choice}.
\newblock In {\em Proceedings of the 30th ACM SIGPLAN Conference on Programming
  Language Design and Implementation}, PLDI '09, pages 38--49, New York, NY,
  USA, 2009. ACM.

\bibitem{Bertolacci-LoopChains}
I.~J. Bertolacci, M.~M. Strout, S.~Guzik, J.~Riley, and C.~Olschanowsky.
\newblock Identifying and scheduling loop chains using directives.
\newblock In {\em 2016 Third Workshop on Accelerator Programming Using
  Directives (WACCPD)}, pages 57--67, Nov 2016.

\bibitem{PLUTO}
Uday Bondhugula, Albert Hartono, J.~Ramanujam, and P.~Sadayappan.
\newblock A practical automatic polyhedral program optimization system.
\newblock In {\em ACM SIGPLAN Conference on Programming Language Design and
  Implementation (PLDI)}, June 2008.

\bibitem{PATUS}
M.~Christen, O.~Schenk, and H.~Burkhart.
\newblock {PATUS: A Code Generation and Autotuning Framework for Parallel
  Iterative Stencil Computations on Modern Microarchitectures}.
\newblock In {\em Parallel Distributed Processing Symposium (IPDPS), 2011 IEEE
  International}, pages 676--687, May 2011.

\bibitem{Hydro2D}
Guillaume~Colin de~Verdi\`{e}re.
\newblock A 2d hydro code for benchmarking purpose, 2017.
\newblock Accesssed September 2017.

\bibitem{COSMO}
Consortium for Small-Scale~Modelling, 2017.
\newblock Accessed September 2017.

\bibitem{Polly}
Tobias Grosser, Armin Groesslinger, and Christian Lengauer.
\newblock {Polly--Performing Polyhedral Optimizations on a Low-level
  Intermediate Representation}.
\newblock {\em Parallel Processing Letters}, 22(04):1250010, 2012.

\bibitem{STELLA}
Tobias Gysi, Carlos Osuna, Oliver Fuhrer, Mauro Bianco, and Thomas~C.
  Schulthess.
\newblock {STELLA: A Domain-specific Tool for Structured Grid Methods in
  Weather and Climate Models}.
\newblock In {\em Proceedings of the International Conference for High
  Performance Computing, Networking, Storage and Analysis}, SC '15, pages
  41:1--41:12, New York, NY, USA, 2015. ACM.

\bibitem{Ng-Contraction}
J.~Ng, Dattatraya Kulkarni, W.~Li, R.~Cox, and S.~Bobholz.
\newblock Inter-procedural loop fusion, array contraction and rotation.
\newblock In {\em 2003 12th International Conference on Parallel Architectures
  and Compilation Techniques}, pages 114--124, Sept 2003.

\bibitem{OpenMP40}
{{OpenMP} Architecture Review Board}.
\newblock {{OpenMP} Application Program Interface Version 4.0}, July 2013.
\newblock Accessed September 2017.

\bibitem{Halide}
Jonathan Ragan-Kelley, Connelly Barnes, Andrew Adams, Sylvain Paris, Fr{\'e}do
  Durand, and Saman Amarasinghe.
\newblock {Halide: A Language and Compiler for Optimizing Parallelism,
  Locality, and Recomputation in Image Processing Pipelines}.
\newblock In {\em Proceedings of the 34th ACM SIGPLAN Conference on Programming
  Language Design and Implementation}, PLDI '13, pages 519--530, New York, NY,
  USA, 2013. ACM.

\bibitem{OPS}
Istv{\'a}n~Z Reguly, Gihan~R Mudalige, Michael~B Giles, Dan Curran, and Simon
  McIntosh-Smith.
\newblock {The OPS Domain Specific Abstraction for Multi-block Structured Grid
  Computations}.
\newblock In {\em Domain-Specific Languages and High-Level Frameworks for High
  Performance Computing (WOLFHPC), 2014 Fourth International Workshop on},
  pages 58--67. IEEE, 2014.

\bibitem{OPS-Fusion}
Istv{\'{a}}n~Z. Reguly, Gihan~R. Mudalige, and Mike~B. Giles.
\newblock Loop tiling in large-scale stencil codes at run-time with {OPS}.
\newblock {\em CoRR}, abs/1704.00693, 2017.

\bibitem{Hydro2D-Pearls}
Jason~D. Sewall and Guillaume~Colin de~Verdi\`{e}re.
\newblock From 'correct' to 'correct \& efficient': A hydro2d case study with
  godunov's scheme.
\newblock In {\em High Performance Parallelism Pearls}, chapter~2, pages 7--42.
  2015.

\bibitem{Pochoir}
Yuan Tang, Rezaul~Alam Chowdhury, Bradley~C. Kuszmaul, Chi-Keung Luk, and
  Charles~E. Leiserson.
\newblock {The Pochoir Stencil Compiler}.
\newblock In {\em Proceedings of the Twenty-third Annual ACM Symposium on
  Parallelism in Algorithms and Architectures}, SPAA '11, pages 117--128, New
  York, NY, USA, 2011. ACM.

\bibitem{YASK}
C.~Yount, J.~Tobin, A.~Breuer, and A.~Duran.
\newblock {YASK--Yet Another Stencil Kernel: A Framework for HPC Stencil
  Code-Generation and Tuning}.
\newblock In {\em 2016 Sixth International Workshop on Domain-Specific
  Languages and High-Level Frameworks for High Performance Computing
  (WOLFHPC)}, pages 30--39, Nov 2016.

\end{thebibliography}

\section{Disclaimers}\label{sec:disclaimers}

\noindent Intel, the Intel logo, Intel Xeon, Intel Xeon Phi and Intel VTune are trademarks of Intel Corporation or its subsidiaries in the U.S. and/or other countries.

Software and workloads used in performance tests may have been optimized for performance only on Intel microprocessors. Performance tests, such as SYSmark and MobileMark, are measured using specific computer systems, components, software, operations and functions. Any change to any of those factors may cause the results to vary. You should consult other information and performance tests to assist you in fully evaluating your contemplated purchases, including the performance of that product when combined with other products.  For more complete information visit  www.intel.com/benchmarks.

Intel does not control or audit third-party benchmark data or the other papers referenced in this document. You should visit the referenced documents and confirm whether referenced data are accurate.

\section*{Acknowledgements}
\noindent  The authors would like to thank the members of the HEAT team at Intel for their support of this project, as well as Guillaume Colin de Verdi\`{e}re of CEA and Grzegorz Kwasniewski, Tobias Gysi, and Torsten Hoefler of ETH Zurich for their help with application codes.

\end{document}